\documentstyle[epsf]{mn}
\def\mmm{(m-M)$_0$}
\def\ebv{E(B$-$V)}
\def\fe{[Fe/H]}

\def\bv{B$-$V}
\def\dv{$\delta$V}
\def\gsim{\;\lower.6ex\hbox{$\sim$}\kern-7.75pt\raise.65ex\hbox{$>$}\;}
\def\lsim{\;\lower.6ex\hbox{$\sim$}\kern-7.75pt\raise.65ex\hbox{$<$}\;}

\title[Be 21]{Old open clusters: the interesting case of Berkeley 21\thanks{
 Based on observations made in La Silla, ESO}}

\author[Tosi et al.]{M. Tosi$^1$, L. Pulone$^{2,3}$, G. Marconi$^2$, 
       A. Bragaglia$^1$ \\
 $^1$ Osservatorio Astronomico di Bologna, Italy, 
      e-mail angela@astbo3.bo.astro.it, tosi@astbo3.bo.astro.it \\
 $^2$ Osservatorio Astronomico Roma -- Monte Porzio, Italy, 
      e-mail pulone@coma.mporzio.astro.it, marconi@coma.mporzio.astro.it\\
 $^3$ European Southern Observatory, Garching bei M\"unchen, Germany,
 lpulone@eso.org}
\date{}

\begin{document}
\maketitle

\begin{abstract}
We present CCD BVI photometry of the old open cluster Berkeley 21, one of
the most distant clusters in the Galactic anticentre direction, and 
possibly the lowest metallicity object in the open clusters sample.
Its position and metal abundance make it very important for the study of the
Galactic disc.

Using the synthetic Colour - Magnitude Diagram method, we estimate values for
distance modulus \mmm = 13.4--13.6,  reddening \ebv = 0.74--0.78 (with
possible differential absorption), and age =  2.2--2.5 Gyr.

\end{abstract}


\section{Introduction}

Old open clusters cover a large range of distances, metallicities, and ages
(Friel 1995), and that warrants their use in investigations of the chemical
and dynamical evolution of our Galaxy. To study  the metallicity and age
distribution of open clusters with Galactocentric distance, and avoid
unnecessary and dangerous biases, a key requisite is homogenous analysis of
very accurate observational data, as discussed by, e.g, Janes \& Phelps (1994,
JP94) Carraro \& Chiosi (1994, CC94), Friel (1995), Twarog et al.
(1997,TAAT97).

This is the fifth paper of a series dedicated to the examination of old open
clusters of different ages and metallicities, and located at different
Galactic radii: for them we measure in a homogenous way distance, age,
reddening and metallicity. These quantities are derived from comparison of the
observed colour-magnitude diagrams (CMDs) to synthetic ones generated by a
numerical code based on stellar evolution tracks and taking into account
theoretical and observational uncertainties (Tosi et al. 1991). These
simulations are much more powerful than the classical isochrone fitting method
to study the evolutionary status of the analysed region and have been
successfully applied both to nearby irregular galaxies (Greggio et al. 1998
and references therein) and to galactic open clusters (NGC2243: Bonifazi et
al. 1990; Cr261: Gozzoli et al. 1996; NGC6253:  Bragaglia et al. 1997;
NGC2506:  Marconi et al. 1997).

Berkeley 21 (Be21) is located toward the Galactic anticentre, at coordinates
RA(1950) = 5:48:42, DEC(1950) = 21:46, and l$_{\rm II}$ =187$^{\circ}$,
b$_{\rm II}$ = $-2.5^{\circ}$. It has already been observed by Christian \&
Janes (1979, hereafter CJ), but their photographic CMD is very shallow, barely
reaching the main sequence Turn-Off (TO). They deduced a substantial 
reddening (\ebv $\simeq$ 1.0), a large distance modulus (\mmm $\simeq$ 16),
and a quite young age ($\sim 10^8$ yr). Much better data have been presented
by Phelps et al. (1994, PJM94) in their compilation of old open clusters,
providing \mmm = 13.9$\pm$0.2 and an age of 2.8 Gyr, derived on the basis of
$\delta V$=1.6 ($\delta V$ being the magnitude difference between TO and clump
stars, JP94). The metallicity has been estimated by medium-resolution
spectroscopy (Friel \& Janes 1993, FJ93), but its actual value strongly
depends on the adopted reddening (\ebv = 0.7$\pm$0.2, Janes 1991), with
[Fe/H]= $-0.97^{+0.3}_{-0.1}$ dex. This large uncertainty, given the fact that
Be21 defines the lowest metallicity limit of the open clusters sample and is
one of the clusters most distant from the Galactic centre, is a further
limitation for studies of the (possible) age and distance relations with
chemical abundance in the Galactic disc (see also Twarog et al. 1997).

In Section 2 we describe the observations and data analysis; in Section 3 we
present the derived CMDs involving BVI photometry and discuss the presence of
binary stars. In Section 4 we compare observed and synthetic CMDs and derive
metallicity, age, distance and reddening. Finally, conclusions will be
reviewed in Section 5.

\begin{table}
\begin{center}
\caption{Log of the observations. The cluster field has its centre at
RA(2000) = 5:51:46, DEC(2000) = +21:48:45. The off-cluster field has
coordinates: RA(2000) = 5:52:08, DEC(2000) = +21:53:53}
\begin{tabular}{lcccccccc}
\hline\hline
\multicolumn{1}{c}{Night} &\multicolumn{1}{c}{Field} &\multicolumn{1}{c}{Tel.} 
 &\multicolumn{3}{c}{Exposure in seconds} 
        &\multicolumn{1}{c}{Seeing} \\
& & & B & V & I & excursion \\
\hline
Mar 4 &Centre &Danish & 120 & 120 & 120 & 0.90-1.20\\
Mar 4 &Centre &Danish &  -  & 120 & 900 & 0.85-1.00\\
Mar 4 &Centre &Danish &  -  & 600 & - & 0.95-1.10\\
Mar 5 &Centre &Danish & 120 & 120 & 120 & 1.00-1.10\\
Mar 5 &Centre &Danish &1500 & 900 &  -  & 0.95-1.10\\
Mar 14 &Ext.  &Dutch & 120  & 60  &  60 & 1.10-1.30\\
Mar 14 &Ext.  &Dutch & 1200 & 480 & 480 & 1.20-1.30\\
Mar 14 &Ext.  &Dutch & 1200 & 480 & 480 & 1.15-1.30\\
\hline
\end{tabular}
\end{center}
\label{tab-log}
\end{table}

\begin{figure*}
\vspace{14cm}
\includegraphics{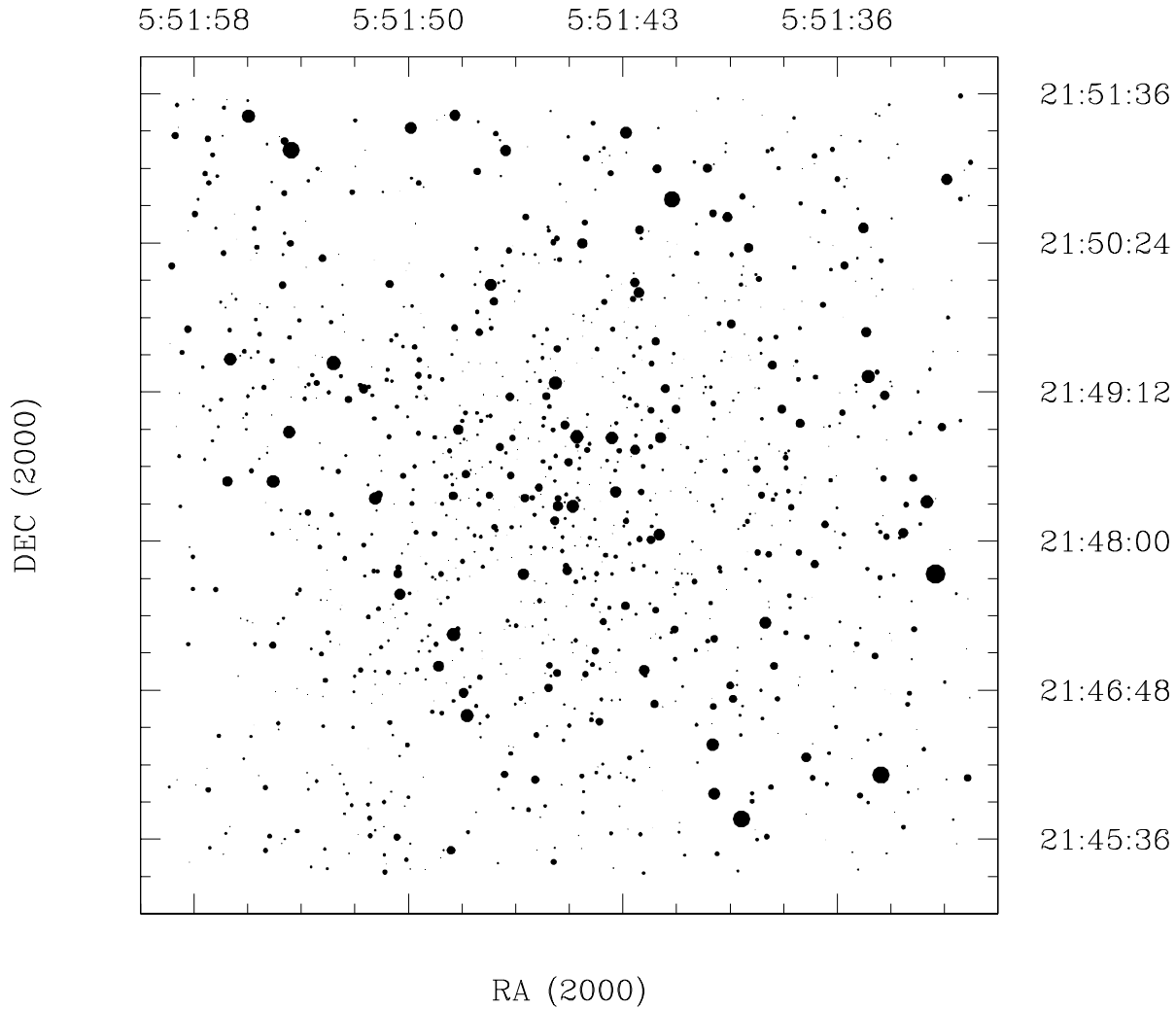}
\caption{Map of the observed field, taken from our $V,B-V$ photometry.
North is up and East left.}
\label{fig-map}
\end{figure*}

\section{Observation and data reductions}

Be21 was observed at the 1.54m Danish telescope located in La Silla, Chile, on
March 4-5, 1995; the field was centered on the cluster. The direct camera
mounted  the CCD \#28, a Tek 1024$\times$1024, with a scale of 0.377
arcsec/pix, yielding a field of view of 6.4$\times$6.4 arcmin$^2$. The
observed region is shown in Fig.~\ref{fig-map}, derived from our photometry
(bright field stars may be missing), and oriented with North up and East left.
At least one of the two nights (March 5) was photometric, while cirri were
occasionally present during the other. Seeing conditions were quite good for
the site/telescope and the seeing excursion for each night is given in Table
1, together with a list of exposures. A second adjacent area, to be used for
field decontamination, was observed with the 0.91m Dutch telescope, also in La
Silla, a few days later. In this case the field of view is of 3.8$\times$3.8
arcmin$^2$.

Standard CCD reduction of bias and dark current subtraction, trimming, and
flatfield correction were performed. We applied to all frames the usual
procedure  for PSF study and fitting available in DAOPHOT--II (Stetson 1992)
in MIDAS  environment. The deepest I frame of each field has been used to
search for stellar objects, setting  the minimum photometric threshold for
object detection at 3 $\sigma$ above the local sky background. The objects
identified in the I band were then fitted in all the others.

The formal errors as given by DAOPHOT are quite small: they range from 0.005
mag for V$<17$ to 0.04 at V$\simeq$22, from 0.001 mag for B$<$18 to 0.08 at
B$\simeq$22, and from 0.005 for I$<17$ to 0.07 al I$\simeq$22. The
circumstance that the synthetic CMDs described in Sect. 4 reproduce quite well
the observed stellar spread in the diagram when adopting these formal errors,
guarantees that even if they do underestimate the actual size of the
photometric errors, the difference must be quite small, thanks to the good
seeing and relatively uncrowded conditions of these data.

Finally, aperture photometry was performed on a few isolated stars (of which 7
were eventually retained) in all images, to compute a correction to the PSF
derived magnitudes and be on the same system as the photometric standard
stars. An indication of the good fit obtained for the PSF of each frame is
that this correction is always small, being of a few hundredths of magnitude
in most fields, and barely reaching 0.1 mag in the very worst cases.

\begin{table}
\begin{center}
\caption{Comparison between our magnitudes and the ones in CJ for the 8 stars
in common. The differences in columns 5 and 6 are in the direction of our
values minus CJ's.}
\begin{tabular}{rccrrr}
\hline\hline
\multicolumn{1}{c}{$N_{CJ}$}       &\multicolumn{1}{c}{$V_{CJ}$} 
&\multicolumn{1}{c}{$(B-V)_{CJ}$}  &\multicolumn{1}{c}{$N$} 
&\multicolumn{1}{r}{$\Delta V$}    &\multicolumn{1}{r}{$\Delta B$}\\
\hline
 250 &12.80  &0.92 &1105 &-0.081 &-0.060\\ 
 254 &13.53  &1.27 &1106 & 0.054 & 0.063\\ 
  14 &13.75  &0.74 &  26 &-0.050 &-0.048\\ 
 260 &13.65  &0.87 &   7 & 0.040 & 0.056\\ 
 211 &15.10  &1.43 &  97 & 0.092 & 0.054\\
  90 &15.32  &0.93 &  93 & 0.022 & 0.015\\
 227 &15.46  &0.88 &  57 & 0.045 & 0.026\\
 168 &16.88  &1.42 & 115 &-0.061 &-0.043\\ 
\hline
\end{tabular}
\end{center}
\label{tab-conf}
\end{table}

\subsection{Photometric calibrations}

The conversion from instrumental magnitudes to the Johnson-Cousins standard
system was obtained using a set of primary calibrators (Landolt 1992) which
spanned a wide range in colour (--0.287 $\leq$ B-V $\leq$ +1.147). Standard
stars fields were analysed using aperture photometry. The calibration
equations were derived using the extinction coefficients for La Silla taken
from the database maintained by the photometric group at the Geneva
Observatory Archive. We obtained equations in the form:
$$ B = b + 0.206 (\pm 0.021) \cdot (b-v) - 4.135 (\pm 0.018) $$
$$ V = v + 0.021 (\pm 0.010) \cdot (b-v) - 3.916 (\pm 0.009) $$
$$ V = v + 0.000 (\pm 0.014) \cdot (v-i) - 4.011 (\pm 0.015) $$
$$ I = i - 0.013 (\pm 0.011) \cdot (v-i) - 4.069 (\pm 0.010) $$
where $b,v,i$ are instrumental magnitudes, while $B,V,I$ are the corresponding
Johnson-Cousins magnitudes. The $V$ values have been obtained from the
equation involving the $b-v$ colour for all stars found and measured in the B
frame (1138 objects), and from the one involving the $v-i$ colour for the
remaining 429 objects. The two calibrations are similar for the stars in
common, without offsets or trends. We calibrated in each filter the March 5
data using these equations and later used them to extend the calibration to
all the other exposures.
   
We have 8 stars in common with the photoelectric observations in CJ, and the
comparison looks good, with no indications of systematic errors in our
photometry: the difference in V magnitude is always less than 0.1 mag, and
even better in B (see Table 2), with a mean difference in both cases of only
0.008 mag. No comparison has been attempted with the photographic values in
CJ, given the very sparse appearance of the diagram.

\begin{table}
\begin{center}
\caption{Completeness of our measurements.
Each value is the average of 5 trials for B, and 3 trials for V and I.
For each magnitude bin, 10 \% of the total number 
of stars were added, following the luminosity function.}
\begin{tabular}{c r r r}
\hline\hline
Mag interv. &\%B &\%V &\%I\\
\hline
$<$ 19.0   & 100 &   100 &  100 \\
19.0-19.5  & 100 &   100 &   95  \\ 
19.5-20.0  & 100 &   100 &   80  \\ 
20.0-20.5  & 100 &   100 &   29  \\ 
20.5-21.0  & 100 &    98 &   10  \\ 
21.0-21.5  & 100 &    78 &       \\ 
21.5-22.0  & 100 &    69 &       \\ 
22.0-22.5  &  89 &    14 &       \\ 
22.5-23.0  &  66 &     2 &       \\ 
23.0-23.5  &  39 &       &       \\ 
23.5-24.0  &  11 &       &       \\ 
24.0-24.5  &   2 &       &       \\ 
\hline
\end{tabular}
\end{center}
\label{tab-compl}
\end{table}

\subsection{Completeness analysis}

We tested the completeness of our luminosity function in the B, V and I band,
using the routine ADDSTAR in DAOPHOT--II. In short, we added to  the original
deepest frames in each filter a pattern of artificial stars, distributed in
colour as the real ones ($\simeq$ $10\%$ of the total in each magnitude bin),
at random positions. The obtained ``artificial frames'' were reduced using
exactly the same procedure and the same PSF used for the original images. We
considered as ``recovered'' only those stars found in their given position and
magnitude bin. The completeness was then derived as the ratio
$N_{recovered}/N_{added}$ of the artificial stars generated. We performed 3 or
5 trials per band, depending on the stability of the results,  and the final
averaged results are reported in Table 3.

\begin{figure*}
\vspace{9cm}
\includegraphics{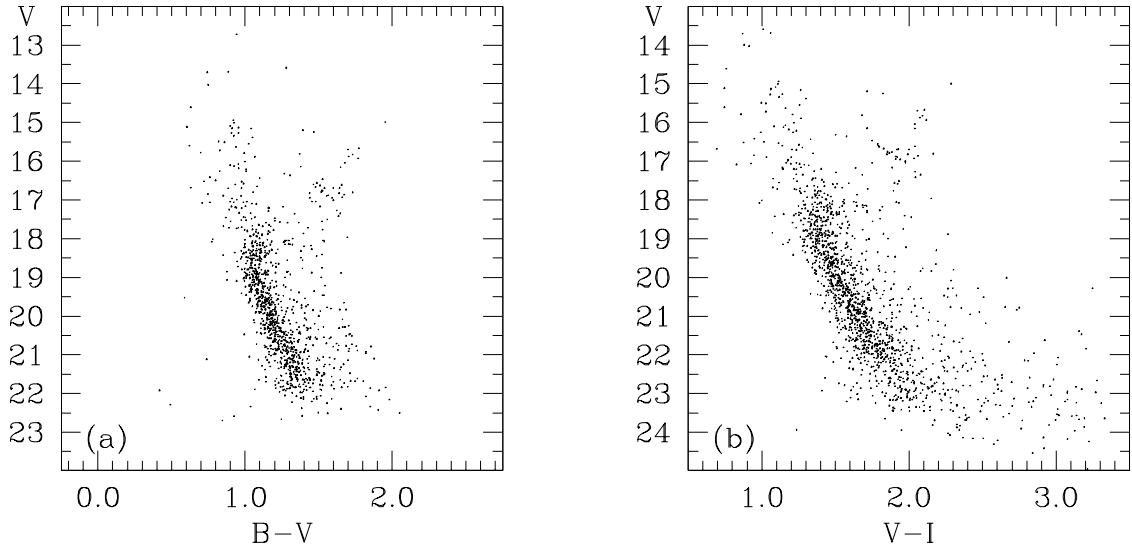}
\caption{Colour magnitude diagrams for our sample: (a) $V,B-V$ CMD, for
1138 stars, (b) $V,V-I$ CMD for 1567 stars}
\label{fig-cmd}
\end{figure*}

\begin{figure}
\vspace{20cm}
\includegraphics{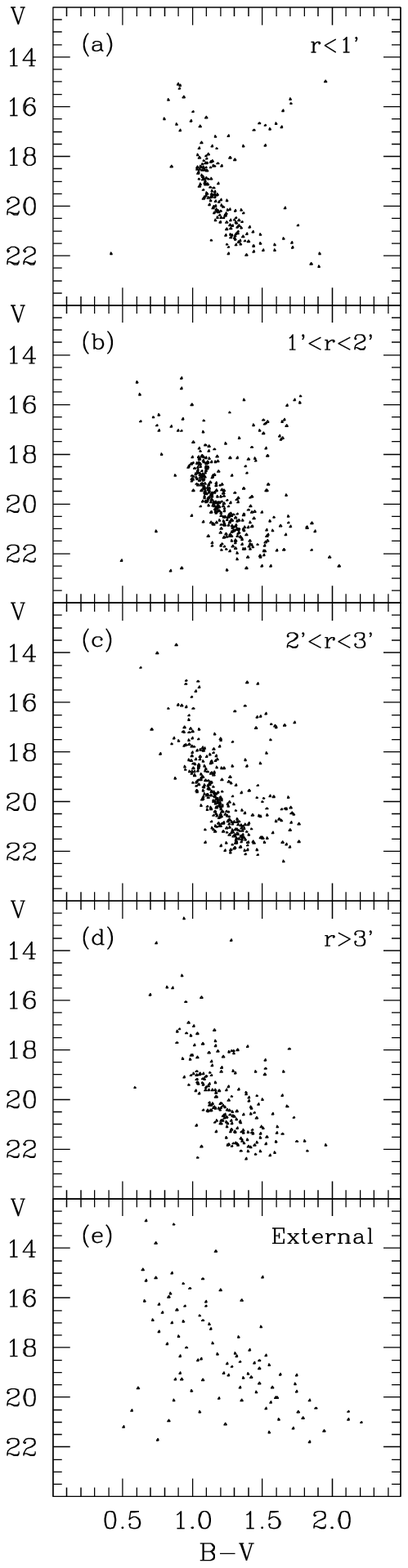}
\caption{(a) to (d) Radial CMD for Be21; (e) external field. Areas are:
1$\pi$, 3$\pi$, 5$\pi$, 4$\pi$, and 4.6$\pi$ respectively }
\label{fig-rad}
\end{figure}

\section{The colour-magnitude diagrams}

The final sample comprises 1567 stars; of them, 1138 have been measured in all
the three bands, while 429 have only V and I values. The complete table with
magnitudes and positions is available electronically from the first author. We
have given positions both in pixel and in equatorial coordinates. The
transformation has been carried out identifying 45 stars well distributed in
the field on the Digitized Sky Survey\footnote{The Digitized Sky Survey was
produced at the Space Telescope Science Institute under U. S. Government grant
NAG W-2166.} images; we are confident of its precision for the central part of
the field, while there might be some distortions (of the order of a few arcsec) 
near the corners.

Fig.~\ref{fig-cmd} shows our results in the $V,B-V$ and $V,V-I$ planes.
The main sequence (MS) appears well delineated in both diagrams, and reaches
more than four magnitudes fainter than the TO.
We estimate for the TO point $V=18.60$, $B-V=1.02$ and $V-I=1.33$.
A red clump of objects, which we attribute to core He-burning stars,
is present at about $V=16.80$, $B-V=1.55$ and $V-I=1.90$.
The subgiant and red giant branches (SGB, RGB) are also visible, although
blurred by field star contamination, considerable especially on the red of the
MS.

Our CMDs (Fig.~\ref{fig-cmd}) are in agreement with those in PJM94 (their 
fig. 12); our MS is, however, narrower and 1.5--2.5 mags deeper, thanks to
the higher quality data.

We plot in Fig.~\ref{fig-rad}(a)-(d) the $V,B-V$ CMDs at different distances 
from the centre (assumed at position X=540, Y=520, or RA(2000)=5:51:45, 
DEC(2000)=+21:48:20); the cluster MS is always visible to the image limit.
The innermost 1 arcmin region (Fig.~\ref{fig-rad}(a)) appears almost
free of field stars, with a narrow MS, and shows presence of blue stragglers
stars (BSS). This is valid also for the adjacent region (Fig.~\ref{fig-rad}(b),
between 1 and 2 arcmin) while the situation is more difficult to disentangle
in the outer rings, where field stars are more numerous.

We note that the BSS seem to be more centrally concentrated than the other 
cluster stars. If we conservatively define as BSS all the objects brighter 
than V=17, and bluer than \bv=1.2, we have 12 of them in the innermost 1 arcmin
(as compared to a total of 192 stars), 12 between 1 and 2 arcmin (total 386
stars, and area 3 times larger) and 17 between 2 and 3 arcmin (total 350 
stars, and area 5 times larger) with increasing relevance of field stars
contamination.
The same appears to be true for binary stars: they are almost as many as
the single stars at the cluster centre, while their relative relevance
decreases outwards (see later).
 
\subsection{Field contamination and differential reddening}

To account for field star contamination, we have observed with the Dutch
telescope a region off-cluster, about 5.5 arcmin North and East of the cluster
centre. The area surveyed by this external field is about 35 \% of that covered 
by the Danish telescope data, and the corresponding $V,B-V$ CMD for the 96
objects found is shown in Fig.~\ref{fig-rad}(e). Taking the area into account,
the number of objects expected to be fore/background stars in the cluster 
$V,(B-V)$ CMD is then about one quarter of the total. Therefore, 
in the inner 1 arcmin cluster members appear to dominate over the field
population: we expect to have only about 20 field stars out of the 
192 (cluster plus field) detected objects.

\begin{figure*}
\vspace{9cm}
\includegraphics{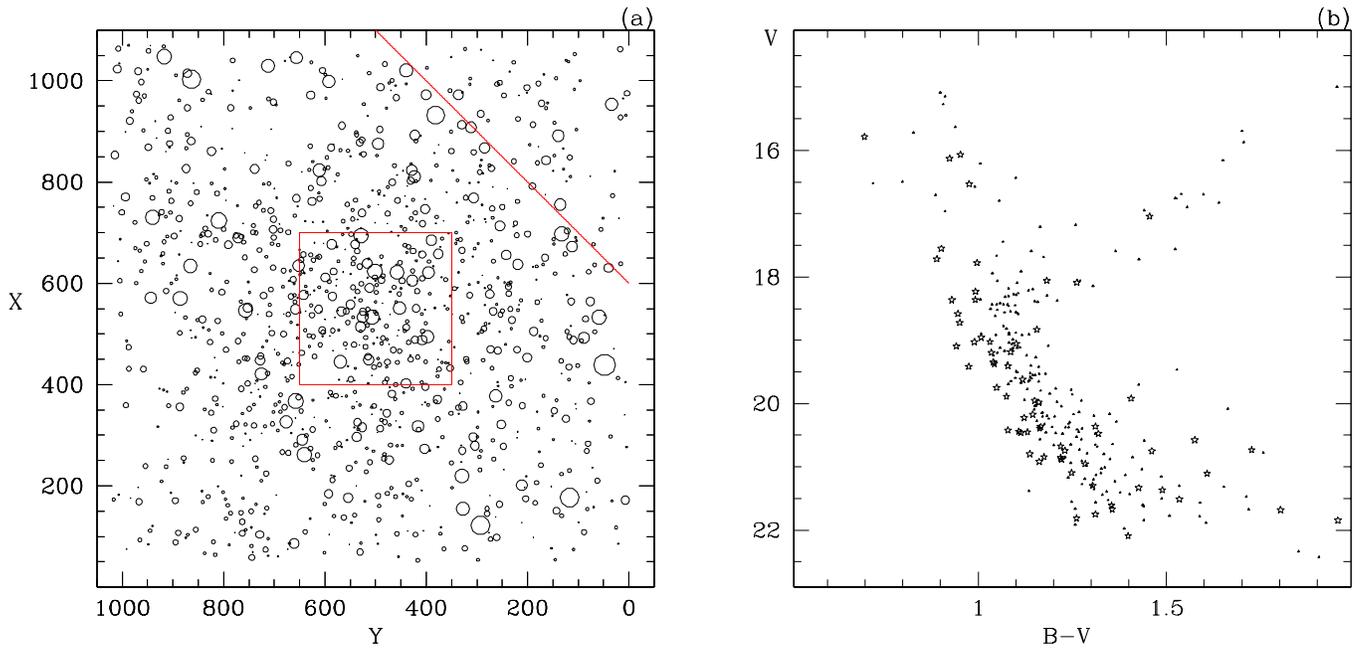}
\caption{(a) Differential reddening is present over the cluster face: in the
upper right corner of our field absorption is less severe. (b) Comparison
between the CMD for the upper right corner (star symbols) and the centre of
the cluster (dots)}
\label{fig-buco}
\end{figure*}

The width of the global MS (Fig.~\ref{fig-cmd}) 
of this cluster appears too large to be 
explained by field stars and photometric errors, unless we pretend to have 
underestimated them by more than a factor of ten, which is quite improbable in 
these photometric conditions. And the presence of a high fraction of binaries 
(see next section) is also insufficient. 
On the other hand, the MS width of the central region (Fig.~\ref{fig-rad}(a))
is instead perfectly reproduceable (see Sect.4) with the estimated photometric
errors when binaries are taken into account.
Since the cluster suffers from large and uncertain absorption (\ebv =
0.7$\pm$0.2, Janes 1991), we have therefore investigated the possibility that 
there  may be differential reddening over its face.
We then divided the field in 9 zones of equal area, compared the resulting CMDs,
and found that one of the corners is less absorbed than the other parts;
the difference in \ebv ~is about 0.1.  By changing position and/or size of the 
zones, we have found that, to a good approximation, we can consider as
less reddened the stars in the North-West corner of our field, as shown in
Fig.~\ref{fig-buco}(a), where also the central box is indicated for comparison.
Fig.~\ref{fig-buco}(b) displays the difference in the CMDs for the
central part and the NW corner, and it is apparent that a significant fraction
of the MS width is due to differential reddening.

\subsection{Binary stars}

\begin{figure*}
\vspace{9cm}
\includegraphics{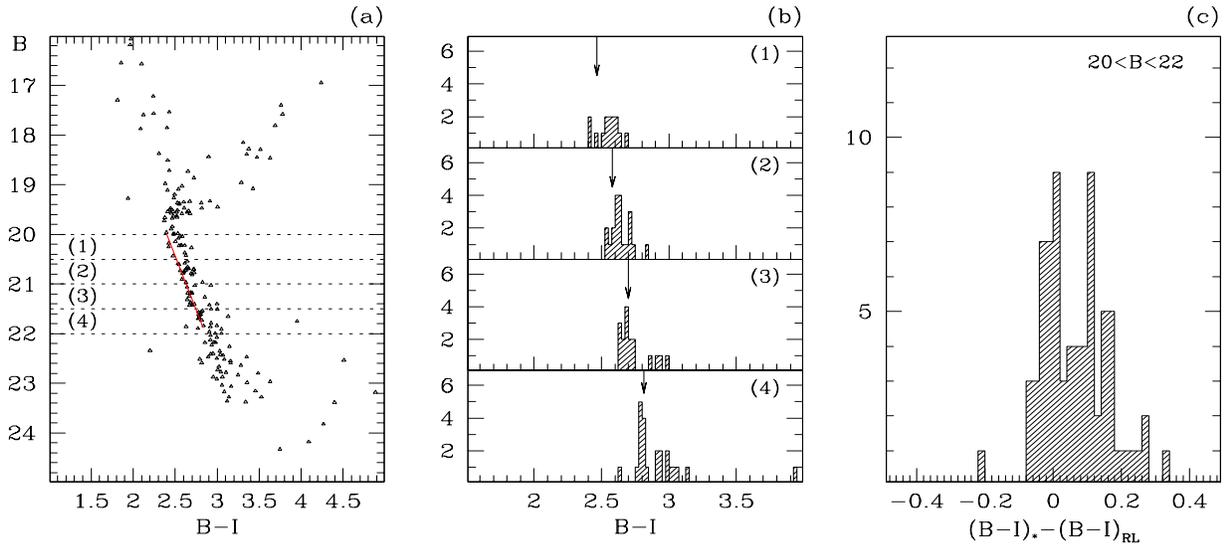}
\caption{(a) $B,B-I$ CMD (the one with the largest baseline in colour possible
with our data). The line is an approximation of the single stars MS ridge
line. (b) Histogram in B-I colour for the four different magnitude intervals.
The arrows indicate the position of the MS "ridge line" in the middle of each
mag bin. (c) Histogram of the difference in colour between stars and the
MS "ridge line" shown in (a).}
\label{fig-bin}
\end{figure*}

At first sight (Fig.~\ref{fig-cmd}) searching for binaries doesn't  look too
promising,  given the width of the MS. If one chooses to work with a larger
baseline in colour (e.g. $B-I$), the situation improves, but significant
results are achieved only when restricting the sample to the very centre of 
the cluster. After a few trials, we decided to concentrate on the inner 1
arcmin, where differential reddening and field contamination are negligible
and the separation between the single and double stars sequences  is
pretty clear (see Fig.~\ref{fig-bin}(a)). We defined a pseudo MS ridge
line (simply joining two points at B about 20 and 22) which fits quite well
the single stars MS. To measure the binary population in a more quantitative
way, we built the histograms in colour of the stars between B=20 and B=22,
divided in 4 bins half a magnitude wide (Fig.~\ref{fig-bin}(b), where the
arrows indicate the position of the "MS ridge line"). 
It is interesting to notice 
that binaries seem to outnumber single stars in at least 2 mag bins. 
To estimate the actual fraction of binaries we have computed the distance of
each star in the interval between B=20 and B=22  from the MS ridge line
(Fig.~\ref{fig-bin}(c)). Two peaks are visible, one located around 
$\Delta(B-I)$=0, which can be considered due to the single stars, and the
second, centered at $\Delta(B-I)$=0.14, due to binary stars. We thus deduce a 
binary fraction of about 45 \%. This estimate is based mostly upon systems of
stars of near equal mass, that show up well separated from the single-stars MS
in the CMD; it is, in this sense, a lower limit to the actual fraction of
binaries.

On the other hand, as said above, binaries are likely to be centrally
concentrated. This is only a rather qualitative statement, but is also 
supported by the central concentration of BSS described above. At the
cluster centre the single-stars and binary sequences are well separated (see
Fig.~\ref{fig-bin}(a)), and there are almost as many objects above and to the
red of the single-stars MS than on the MS itself. The situation changes in the
adjoining ring, where the picture is blurred by field star contamination, but 
there appears to be a clustering of stars to the blue of the MS, with decreasing
density to the red. The picture is consistent with mass segregation of the more
massive objects (MS binaries, BSSs) toward the centre of the cluster.

\begin{table*}
\begin{center}
\caption{Stellar evolutionary tracks adopted for the synthetic CMDs}
\begin{tabular}{lccccll}
\hline\hline
Model & Y & Z & M$_{min}$    & M$_{max}$    & Reference & Notes\\
      &   &   &(M$_{\odot}$) &(M$_{\odot}$) &           & \\
\hline
FRANEC & 0.27 & 0.006 & 0.6 & 9   & Castellani et al. 1993 
                                                      & to AGB-tip, LAOL op.\\
FRANEC98 & 0.23 & 0.001 & 0.7 & 9   & Straniero, priv.comm.
                                                      & to AGB-tip, OPAL op.\\
FRANEC98 & 0.26 & 0.006 & 0.7 & 9   & Straniero, priv.comm.
                                                      & to AGB-tip, OPAL op.\\
FRANEC98 & 0.27 & 0.010 & 0.7 & 9   & Straniero, priv.comm.
                                                      & to AGB-tip, OPAL op.\\
Geneva & 0.24 & 0.004 & 0.8 & 120 & Charbonnel et al. 1993 & only to RGB-tip\\ 
Geneva & 0.24 & 0.001 & 0.8 & 120 & Schaller et al. 1992 
                                          &Charbonnel et al. 1996 for e-AGB \\
Padova & 0.24 & 0.004 & 0.6 & 120 & Fagotto et al. 1994  & to AGB-tip\\
Padova & 0.25 & 0.008 & 0.7 & 120 & Alongi et al. 1992 & to AGB-tip\\
\hline
\end{tabular}
\end{center}
\label{tab-mod}
\end{table*}

\section{CLUSTER PARAMETERS}

To derive the values of the cluster parameters, we have applied to Be21 the
approach of CMD simulations described by Tosi et al. (1991) and already
employed for four other old open clusters. Due to the existing estimate of
the cluster metal content (FJ93), we have restricted the sample of stellar 
evolutionary tracks adopted to create the synthetic CMDs to the sets with
metallicities near that value (i.e. Z $\simeq$ 0.002). These sets 
are summarized in Table 4. For each set of stellar models,
we have performed several MonteCarlo simulations for any reasonable
combination of age, reddening and distance modulus.

The incompleteness factors and the photometric errors in each magnitude bin
assigned to the synthetic stars in each photometric band are those derived
from the observed data and described in Sect. 2. 

Generally speaking, the CMD of Be21 is not easily reproduceable by stellar
models. This is  due to the combined effects of field
contamination and differential reddening, to the presence of a large number 
of blue stragglers candidates and to the underpopulation and/or short 
extension of the subgiant branch as compared to that expected for a system
with the magnitude difference between clump and turnoff stars observed in Be21.
To minimize at least some of these problems in the 
comparison between observed and simulated CMDs, we have restricted the complete
analysis to the inner cluster region where field contamination and reddening
variations are not severe. 
For sake of consistency, we have also made some
simulations of the whole cluster, as well as of different sub-regions and 
always found results equivalent to those described below.

The region of the central 1 arcmin contains 192 stars, of which 12, with
V$\leq$17 and B--V$\leq$1.2, are either BSSs or fore/background objects and
cannot be predicted in the synthetic diagrams. By scaling the number of stars
detected in the external field (Fig.~\ref{fig-rad}(e)) to the area enclosed in
the central 1 arcmin, we can also estimate that 21 of the 192 stars are
presumably non cluster members, preferentially distributed along a wide
sequence roughly parallel to the MS and overlapping the BSS region. Since the
synthetic CMD cannot predict either BSSs or non-member objects, to reproduce
the same number of stars as the probable cluster members of the central
region, our simulations contain 159 objects.

From the point of view of comparing the CMD morphology, the external stars are
few and sparse and therefore only moderately relevant. They may instead affect
the luminosity function (LF). To check this effect, we have performed several
random selections of the 21/96 stars in the external field and found that
their distribution invariably looks sparse and diffuse. At any rate, to take
them into account, in the following we compare the observed LF of the central
cluster region both to the pure synthetic LFs and to those resulting from the
sum of the synthetic stars plus the 12 BSSs and the 21 randomly selected
external objects.

Since the data show evidence for a large fraction of binary stars among the 
cluster members, we have included them in the synthetic diagrams. To do this,
a mass ratio has been associated to each system via random extractions from a
flat distribution. As for the other clusters, we have followed the 
prescriptions given by Maeder (1974) to attribute colours and magnitudes to 
systems with different primary/secondary mass ratios (see Bragaglia et al. 
1997 for details). 
All the stellar models lead to synthetic CMDs in better agreement with the 
data when the assumed fraction of binaries is 45$\%$; a result in striking
agreement with the fraction empirically derived in Sect. 3.2

\begin{figure*}
\vspace{17cm}
\includegraphics{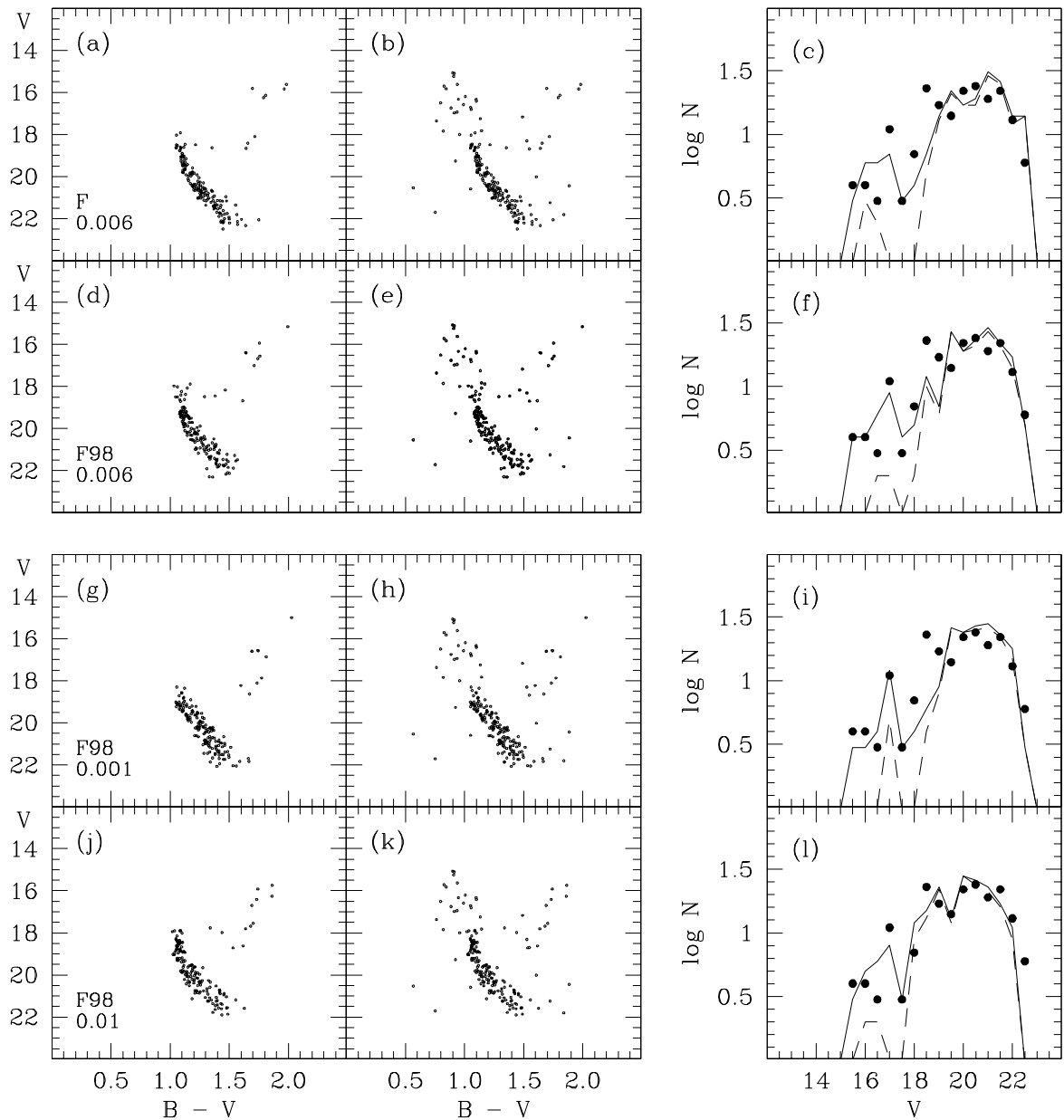}
\caption{Models in better agreement with the data, for each set of FRANEC 
tracks (see Tables 4 and 5).
Each row shows the synthetic CMD on the left, its superposition to the 
CMD of 12 BSSs and 21 external objects (see text for details) in the middle, 
and the comparison of the two corresponding LFs on the right (full dots for 
observational data points, dashed line for pure synthetic ones, solid line for 
the latter plus BSSs and external objects). The metallicity is indicated in the 
bottom left corner of each row. }
\label{fig-sim1}
\end{figure*}

\begin{figure*}
\vspace{17cm}
\includegraphics{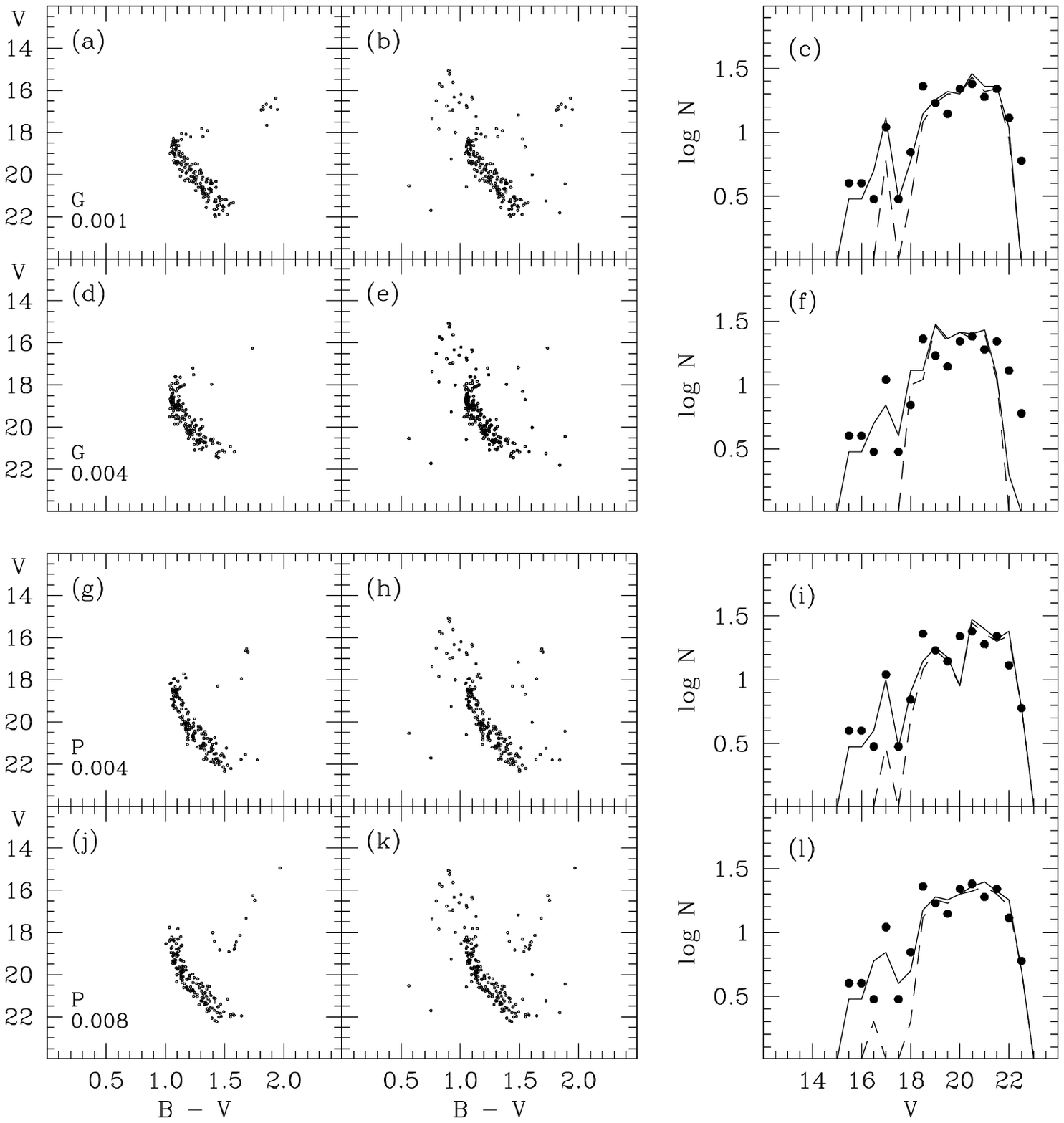}
\caption{Models in better agreement with the data, for the Geneva (two
top rows) and Padova (two bottom rows) sets of tracks (see Tables 4 and 5).
Each row shows the synthetic CMD on the left, its superposition to the 
CMD of 12 BSSs and 21 external objects (see text for details) in the middle, 
and the comparison of the two corresponding LFs on the right (full dots for 
observational data points, dashed line for pure synthetic ones, solid line for 
the latter plus BSSs and external objects). The metallicity is indicated in the 
bottom left corner of each row. }
\label{fig-sim2}
\end{figure*}

\subsection{Results with FRANEC stellar models}

The FRANEC stellar tracks have been recently updated to improve the input
physics and include the OPAL opacities (Straniero et al. 1998). Thanks to
their authors (Straniero, 1997 private communication), we have been able to
use their new version in advance of publication. Generally speaking, the new
models show slightly fainter MS and shorter blue loops (i.e. core He-burning
phases in intermediate mass stars) than the older ones. For sake of
homogeneity with our previous studies, based on the 1993 FRANEC tracks, we have
simulated the CMD of Be21 also with the old set at Z=0.006.

The most representative results obtained with the FRANEC stellar models are
displayed in Fig.~\ref{fig-sim1}, where each row corresponds to the best case
resulting from the set of FRANEC tracks indicated in the bottom left corner of
that row. The age, reddening and distance modulus corresponding to such cases
are listed in Table 5. The three panels in each row show: i) the pure synthetic
CMD with 159 stars, on the left; ii) the sum of the latter with the observed
CMD of the 12 BSSs candidates and of the 21 probable external objects in the
central 1 arcmin (i.e. a total of 192 objects), in the middle; iii) the
luminosity functions of the synthetic objects (159, dashed line), of the sum
of these stars plus the BSSs and external objects (192, solid line), and of
192 the stars measured in the central 1 arcmin (full dots), on the right.

Panel (d) of Fig.~\ref{fig-sim1} shows one of the synthetic CMDs in better
agreement with the data. It is based on the FRANEC98 tracks with Z=0.006 and
corresponds to: $\tau$=2.5 Gyr, \mmm=13.6 and \ebv=0.74. The two latter values
are roughly consistent with the \mmm=13.9$\pm$0.2 and the \ebv=0.7$\pm$0.2
estimated by PJM94 and Janes (1991), respectively. The age is a bit younger,
but it is rather difficult to reproduce the cluster features with their value
of 2.8 Gyr: the synthetic clump would be too bright, since the older the
system, the larger the magnitude difference between TO and clump stars. In
fact, with these tracks we obtain a fit of similar quality if we assume an age
as young as $\tau$=2.3 Gyr (with \mmm=13.7 and \ebv=0.74).

The LF corresponding to this synthetic CMD is represented by the dashed line
in panel (f) of Fig.~\ref{fig-sim1}. It shows an apparent paucity of bright
stars. However, when the BSSs and the appropriate fraction of field
contaminating stars are included (solid line), the resulting LF is consistent
with the empirical one (full dots). Notice that we are examining the very
central region of the cluster, where paucity of low mass stars with respect to
higher mass ones, produced by dynamical evolution and mass segregation, should
be in principle more noticeable. In the case of Be21, however, not many of the
low mass stars may have been forced to move to the outermost regions. In fact,
the agreement between synthetic and empirical LF at the faintest mag bins
shows that Be21 has not lost an appreciable fraction of its lowest mass stars,
contrary to other clusters of similar age, but located in inner Galactic
regions (e.g. NGC6253, Bragaglia et al. 1997).

The top row in Fig.~\ref{fig-sim1} shows the best case resulting from the old
FRANEC tracks with Z=0.006. It assumes $\tau$=2.3 Gyr, \mmm=13.7 and
\ebv=0.78, i.e., similar age and distance as with the new FRANEC tracks, but
slightly higher reddening. The reddening difference is due to the fact that
the old tracks are slightly bluer than the new ones, possibly because of the
different adopted opacities and input physics that somehow mimic a lower
metallicity (Cassisi et al. 1994). The SGB shows here a larger colour
extension (leading also to a too red position of the RGB) than with the
FRANEC98 models, and this is probably related to the above effects as well.

Since spectroscopic measurements of stars in Be21 (FJ93) have provided
[Fe/H]=--0.97, corresponding to Z$\simeq$0.002, one would expect to obtain a
better reproduction of the observed CMD and LF by adopting the set of tracks
with lower metallicity, Z=0.001. However, this is not the case for Be21, due
to the fact that lower metallicity models have bluer MS and more extended SGB.
The intrinsically bluer MS needs fairly high values
(0.8$\lsim$E(B-V)$\lsim$1.0) for the reddening, to properly fit the
observational MS. These reddenings are roughly consistent with that derived by
Janes (1991), but, when combined with the large SGB extension, imply RGBs much
redder than observed. These problems with the SGB and RGB make the synthetic
CMDs with Z=0.001 in poor agreement with the data, unless a quite old age
($\sim$ 4 Gyr) is assumed for the cluster, in which case the SGB obviously
shrinks and the RGB goes back to the appropriate position. Unfortunately, such
old ages inevitably imply clumps much brighter than observed and are therefore
unacceptable. In the third row of Fig.~\ref{fig-sim1} we show the best
compromise for CMD (and LF) obtained with the FRANEC98 tracks with Z=0.001. It
assumes $\tau$=2.8 Gyr, \mmm=13.4 and \ebv=0.93. Notice that, in addition to
the above problems, the straight shape of the bright MS does not reproduce
satisfactorily the observed MS either. We thus exclude that this set of
stellar models be appropriate for Be21.

To further verify what is the extent of the metallicity effect on the cluster
parameters derived with this method, we have also simulated the observed CMD
with the FRANEC98 tracks with Z=0.01, obtained (Straniero 1997, private
communication) by interpolating between Z=0.006 and Z=0.02. This metallicity
is admittedly too high, when compared with that spectroscopically derived, but
it has been found in several other occasions (e.g. Cassisi et al. 1994,
Greggio et al. 1998) that the nominal metallicity of stellar models does not
always coincides with the actual one. We thus consider it worthy to use also
this set, despite its formally high Z. Unfortunately, the colour extension of
the SGB remains too large while the luminosity of the RGB clump becomes
significantly higher (probably as a consequence of the higher helium content)
and we are forced to select models with quite younger ages to keep the clump
at a magnitude level consistent with the data. The bottom row of
Fig.~\ref{fig-sim1} shows the best case with Z=0.01, corresponding to
$\tau$=1.5 Gyr, \mmm=13.7 and \ebv=0.76.

\begin{table}
\begin{center}
\caption{Summary of best-fits results for distance modulus, 
age, and reddening for each set of stellar models. The FRANEC98 Z=0.01
models are interpolated between Z=0.006 and Z=0.02}
\begin{tabular}{lrccc}
\hline\hline
\multicolumn{1}{c}{Model} &\multicolumn{1}{c}{Z}
&\multicolumn{1}{c}{(m-M)$_0$}
&\multicolumn{1}{c}{$\tau$ (Gyr)} &\multicolumn{1}{c}{\ebv} \\
\hline
FRANEC   & 0.006 & 13.7 & 2.3 & 0.74  \\
FRANEC98 & 0.001 & 13.4 & 2.8 & 0.93  \\
FRANEC98 & 0.006 & 13.6 & 2.5 & 0.74  \\
FRANEC98 & 0.010 & 13.7 & 1.5 & 0.76  \\
Geneva   & 0.001 & 13.4 & 2.1 & 0.97  \\
Geneva   & 0.004 & 13.4 & 2.1 & 0.60  \\
Padova   & 0.004 & 13.5 & 2.2 & 0.78  \\
Padova   & 0.008 & 13.7 & 2.3 & 0.70  \\
\hline
\end{tabular}
\end{center}
\label{tab-ris}
\end{table}

\subsection{Results with Geneva stellar models}

The Geneva tracks presumably most appropriate for the metallicity of Be21 are
those with Z=0.001 and with Z=0.004. Unfortunately the core-He-burning phase
of low mass stars, corresponding to the clump, is not available for the
latter, and no direct comparison with this observed feature is thus feasible.

The second row of Fig.~\ref{fig-sim2} shows one of the synthetic CMDs 
with Z=0.004 in better agreement with the data, although definitely not 
satisfactory. It corresponds to $\tau$=2.1 Gyr, \mmm=13.4 and \ebv=0.6. 
The lack of He-burning low-mass models, leading to the absence of synthetic 
stars
in the clump phase, prevents any further check on the self-consistency of
these values. However, the shape of the MS (which is quite round in this case,
but becomes too straight for younger ages and too curved for older ones)
and the relative number of objects in the various evolutionary phases limit 
quite strongly the possible range
of ages attributable to Be21 with this set of tracks and, consequently, the
corresponding reddening and distance. In other words, even without the clump,
we are confident that the cluster parameters derived with this set of models
cannot differ significantly from these values.

The Geneva models with Z=0.001 do have the clump,  allowing more strict
constraints on the distance of the cluster. Besides, their metallicity is
nominally closer to that spectroscopically determined for the cluster.
Nonetheless, the fit to the observational data is much poorer than with the
Z=0.004 tracks, due to the excessive colour extent of the subgiant branch,
usually attributable to an insufficient metal content. 
In the best (less bad) case, shown in the top row of Fig.~\ref{fig-sim2}, 
we derive: $\tau$=2.1 Gyr, \mmm=13.4 and \ebv=0.97. 
The inadequacy of these tracks is also apparent from the very high reddening 
required to compensate the excessive blueness of their MS.
As already mentioned for the FRANEC models, this problem may be attributed 
to an insufficient metallicity but,
possibly, also to details in the input physics which may mimic a different
metal content. This latter possibility is suggested by the circumstance that
the same set of Geneva stellar models with Z=0.001 reproduces quite well the
CMD and the LF of the stars in the post-starburst galaxy NGC 1569 
(Greggio et al. 1998), whose metallicity, as derived from HII region 
spectroscopic measurements, is Z=0.004. Despite the different mass range of
the stars visible in the two systems (massive stars in NGC 1569 and low mass 
stars in Be 21), we consider it rather improbable that the same set of tracks 
be simultaneously too metal rich in the massive star range and too metal poor
in the low mass range.

\subsection{Results with Padova stellar models}

The third row of Fig.~\ref{fig-sim2}, 
shows one of the best synthetic CMD resulting from the Padova tracks
with Z=0.004; it provides $\tau$=2.2 Gyr, \mmm=13.5 and \ebv=0.78. The 
corresponding LF, once implemented with the BSSs and the appropriate fraction 
of field contaminating stars, is in agreement with the data, showing that
the relative timescales of the various evolutionary phases are well
reproduced by these tracks and that this cluster has not suffered any 
significant loss of stars. This is possibly the best model out of all our
simulations based on any stellar evolution set.

Since the cluster features are all rather well reproduced by these tracks,
the choice of the age (and therefore of the distance and reddening) is in
this case more restrictive than with the other sets, due to the 
characteristics of the clump luminosity and of the SGB extension.
Had we assumed a slightly younger age 
(e.g. 2 Gyr), the SGB would have turned out too extended 
in colour and the RGB would have been much too red. Had we instead assumed a 
slightly older age (e.g. 2.3 Gyr), the clump would have been too bright, due 
to the larger magnitude difference between TO and clump. For ages older than 
2.4 Gyr, also the fit to the MS worsens because its shape becomes too round
with respect to the observed one.

It is worth emphasizing that these best fitting models with Z=0.004, provide
age and distance in agreement with the values derived from the Geneva tracks
of same Z (see Table 5). The two derived reddenings are instead quite
different. This is due to actual discrepancies in the 
predictions of the adopted stellar models, and cannot be ascribed to 
differences in the transformation from the theoretical to the observational
plane. In fact, contrary to what is usually done, our procedure consists in
adopting the theoretical stellar evolution tracks and transform their log L
and log T$_e$ into M$_V$ and B--V by using for all sets the same conversion
tables. The possible discrepancies between different stellar
tracks explain why we consider of fundamental importance to always compare
the observational CMDs with more than one set of stellar models, to have at 
least an estimate of the corresponding uncertainties.

We have also created synthetic CMDs based on the Padova models with Z=0.008,
but their fits are worse than those with Z=0.004. Rather, with Z=0.008
the SGB is almost as extended in colour and much more populated (i.e. with 
longer relative lifetimes) and, therefore, less consistent with the observed
CMD of Be21. This is visible in the bottom row of Fig.~\ref{fig-sim2} where
the model with $\tau$=2.3 Gyr, \mmm=13.7 and \ebv=0.7 is shown: the SGB and
faint RGB are excessively populated.

Using Padova tracks with metallicity lower than Z=0.004, we would find again 
SGBs with excessive colour extents and quite high required reddenings, as 
already found for the other sets from FRANEC and Geneva.

\begin{table*}
\begin{center}
\caption{Comparison of ages for our sample of old clusters derived: by
ourselves (col. 2), from the MAI (JP94, col. 4, using \dv ~in col. 3) and
from CC94 (or Carraro et al. 1998 when indicated, 
using either the $\Delta$V in col. 5 or direct analysis). 
The MAR is given in col. 9 and is derived from $\delta V_T$ and
$\delta(B-V)_T$ (col. 7 and 8); the actual age value depends on the 
adopted calibration.
Also given are the parameters for Be17, considered the oldest open cluster of
our Galaxy (Phelps 1997).}
\begin{tabular}{lccccccccl}
\hline\hline
Cluster & $\tau$ & $\delta$V & MAI &$\Delta$V & CC94 &$\delta V_T$ 
                                                 &$\delta(B-V)_T$ &MAR &Notes\\
        & (Gyr)  &  (mag)    &(Gyr) &  (mag)   & (Gyr) & (mag) & & &\\
\hline
Be21    &2.2 &1.8 &2.8 &     &3.1 &1.20 &0.62 &1.94 &Carraro et al. 1998\\
Cr261   &7.0 &2.6 &9.5 &     &7.0 &2.56 &0.73 &3.51 &\\
NGC2243 &5.0 &2.2 &5.6 &2.15 &4.5 &2.00 &0.62 &3.61 
                                           &red clump: used brighter part\\
NGC6253 &3.0 &2.0 &4.4 &     &    &1.80 &0.63 &2.86 &\\
NGC2506 &1.6 &1.5 &2.5 &1.75 &1.9 &1.20 &0.68 &1.76 &\\
&&&&&&&\\
Be17    &12$^{+1}_{-2}$ &2.7-2.8 &10.9-12.6 & &9.0 & & & 
                                            & Carraro et al. 1998\\
\hline
\end{tabular}
\label{tab-mar}
\end{center}
\end{table*}

\section{Summary and Discussion}

Although the fits of observed and simulated diagrams are not as satisfying for
Be21 as they have been for other systems examined with the same method, 
we have been able to
determine a fairly consistent confidence interval for its distance, age,
reddening and metallicity (see Table 5) by selecting the most reliable among
the models described in the previous section.
They place it among the old metal poor open clusters, in a
region far from the Galactic centre and of moderately high reddening.

\subsection{Distance and reddening}

We have derived a distance slightly smaller than previous studies, 
while our evaluation of the reddening is fairly consistent with past
works. No previous indication of differential reddening was given in
the literature, but our data definitely show it.

JP94 found, for the red clumps of 23 open clusters with \dv$\ge$1.0
(i.e. older than about 1.5 Gyr) a  mean absolute magnitude $M_V=0.9 \pm 0.4$, 
and a mean intrinsic colour  $(B-V) = 0.95 \pm 0.10$.
In our case, these mean values, when applied to the observed
$V$=16.80 and $B-V$=1.55, would imply
$(m-M)_V$=15.90, and \ebv=0.60, or $(m-M)_0$=14.04.
From our best simulations, we obtain instead $(m-M)_0 \simeq $ 13.5, \ebv=0.76,
corresponding to $(m-M)_V$=15.86.
In other words, the clump-based distances seem to agree, but the colour of the
clump stars seems to be quite different from the mean.
Part of this discrepancy may be due to the high reddening affecting Be21,
whose differential effect on blue and red stars leads to an apparent shrink
by 0.04 of the true colour difference between TO and clump stars (Fernie 1963,
Twarog 1998 private communication).
On the other side, TAAT97, derive a  mean  $M_V=0.6 \pm 0.1$ for ten clusters
not too metal-rich ranging from NGC7789 to Mel66, i.e. approximately from 1 to
5 Gyr, for which they try to measure the distance in a quite homogenous way.
This translates, in our case, to $(m-M)_V$=16.2; since 
they do not cite a mean intrinsic $(B-V)$ no further comparison with our
best choice for the reddening is possible.

There are no completely reliable reddening determination for this cluster
since the UBV data of CJ do not reach the MS, but our determination and that
by Janes (1991) agree well. We have further compared our finding with what
is expected from the spatial distribution of interstellar extinction near the
Galactic plane. To this end we have considered the studies of FitzGerald
(1968, fig. 3h) and Neckel \& Klare (1980, fig. 6i). In both cases, a
reasonable estimate deduced from their  data for low Galactic latitudes and
the right longitude, is \ebv $\sim$ 0.8; FitzGerald's (1968)
observations also allow for a lower value, closer to 0.5,
but seem to exclude the high values, close to 1, needed by the lower 
metallicity tracks of any group.

Janes (1991) and JPM94  give a distance modulus \mmm=13.9$\pm$0.2,  somewhat
larger than our results for every set of tracks.
Carraro et al. (1998), working on the same data, cite a
Galactocentric distance of 14.5 Kpc, also implying a distance modulus
\mmm$\simeq$13.9.
We have no good explanation for this difference, but we must emphasize that 
adopting \mmm=13.9 we would be forced to select younger ages, and this would
have two major drawbacks: a worse disagreement with literature ages (see next),
and a worse reproduction of the MS shape in the synthetic CMDs.

\subsection{Age}

Also in the case of the age, we seem to have found a value lower than given in
literature. We can explain the discrepancy partly  by the different techniques
adopted, partly by the better quality of our data. The
various parameter combinations all converge to a fairly small range of
possible ages (2.1 to 2.8 Gyr, with favorite age around 2.2 Gyr). In fact, the
only largely discrepant value found in our analysis is for the FRANEC98 Z=0.01
tracks, a value in strong disagreement with the spectroscopically determined
metallicity. Despite the uncertainties involved in the age determination
with our method, we consider it still more reliable than ages derived by other
means.

Nonetheless, it is not always feasible to determine the age of a cluster 
with the proper method of synthetic diagram fitting: to do so, high
quality data, both deep and precise, are needed, and the process itself is
complex. To apply this technique to all the objects of interest takes a
long time, while the properties of the whole sample of open clusters are
needed to study the Disc population and evolution. For this reason, several
parameterizations of cluster ages,  based on a handful of well studied
objects, have been proposed. These methods, if uncertain in absolute value for
the single cluster, yield a reasonable age ranking for the cluster system.

These parameterizations are usually based on a difference, in magnitude and/or
in colour, between well recognizable points of the CMD (usually the TO and the
red clump), as this is much easier to measure than any absolute quantity.
Note though that the precise definition of the two points, and especially
of the TO, changes among authors. We will cite here the three following
examples:
i) Anthony-Twarog \& Twarog (1985 and later works) use the magnitude
difference between the red giant clump and {\it the brightest point at the TO}
($\delta V_T$) coupled with the difference in colour between the red giant
branch at the position of the clump and the bluest point of the TO
($\delta(B-V)_T$); ii) JP94 use a similar \dv, but measured between the red
clump and {\it the inflection point between the MS TO and the base of the
giant branch}; iii) CC94 define their $\Delta V$ as JP94, but assume
that {\it the reference TO luminosity is 0.25 mag fainter than observed in the
CMD}, to take into account the fact that presence of unresolved binaries tends
to brighten the TO region.
In the case of Be21, we have: \dv=1.8 (our measure), 
$\delta V_T$=1.2, \dv(JP94)=1.6, $\Delta V$=1.55.
All these different definitions try to circumvent the problem that the TO
point is not always easily identified in open clusters, due to field stars
and binaries contamination and/or paucity of stars. The strength of our kind of
analysis is that we do not judge on the basis of the observed CMD alone:
we know from the tracks the exact location of the TO in each of our
simulated CMDs. Given this, we have chosen to measure the magnitude difference
as defined above at what we believe to be the true TO, i.e. at the 
point corresponding to the hottest MS point in the evolutionary tracks.

JP94 correlated \dv ~with cluster ages. The calibration of their Morphological
Age Index (MAI, expressed in Gyr) translates for Be21 to an age of 2.8 Gyr
(based upon \dv=1.6, PJM94), marginally inconsistent with what we get from
the direct comparison with evolutionary tracks (see Section 4). 
The age difference does not arise from any discrepancy in the two sets
of data: we find  \dv=1.8, quite consistent with the value given by PJM94
considering that we measure it in a slightly different way.
However, we have found in the past (e.g. in the case of NGC2506, Marconi 
et al. 1997) that the MAI tends to overestimate ages.

Anthony-Twarog \& Twarog (1985, revised by Twarog \& Anthony-Twarog 1989)
proposed the so called Morphological Age Ratio (MAR), defined as 
MAR = $\delta V_T / \delta (B-V)_T$ (see above).
This index is independent of reddening and almost independent of metallicity
(Anthony-Twarog \& Twarog 1985, Buonanno et al. 1989).
The calibration of
the relation between MAR and ages has changed from: age = 1.4$\times$MAR Gyr
(Anthony-Twarog \& Twarog 1985) to: age = 2.0$\times$MAR Gyr (Twarog \&
Anthony-Twarog 1989). 
Applying their definition to our CMD, we find the values given in Table 6, 
and an age of about 2.7 to 3.9 Gyr, depending on
the adopted calibration.
With no attempt to give a new calibration of the MAR-age relation,
simply adopting for the clusters we studied the parameters and ages in Table 6,
we find:  age = 2.3 $\times$ MAR -- 2.6 ($r.m.s.$ = 0.9). 
We did not include Be17 in this
computation: it represents an extreme of the interpolation and we felt that
the parameters derived from the published diagrams were too insecure.
This relation gives for Be21 an age of 1.9 Gyr.

Carraro et al. (1998) derive for Be21 an age of 3.1 Gyr, based on the PJM94
data and the synthetic CMD method using the Padova tracks. The difference with
our results, obtained employing the same sets of tracks (although we do not
interpolate in metallicity as they do), may perhaps be explained simply with
the worse quality of the observational data they use. Certainly, in no case
are we able to reach self-consistently such a large age.

\subsection{Metallicity}

We note that our method is unable to solve the problem of the cluster precise
metal abundance. The comparison with evolutionary tracks can only give a
coarse indication of metallicity. Too many variables are present in tracks
computation to discriminate metallicity to such an extent.
In fact, tracks nominally closer to the metallicity derived for Be21 from
spectroscopic measurements ([Fe/H]=--0.97, or Z $\simeq$ 0.002, FJ93) appear
less consistent with the observed CMD than tracks more metal rich, because of
the large colour extent of the subgiant branch and, in some cases, of an
excessively high reddening required to reproduce the observed colour of the
MS.

Anyway, what can be said is that the best fits are obtained for the slightly
more metal-rich combinations, i.e. for Z=0.006 (\fe $\simeq$ -0.5) or 0.004 
(\fe $\simeq$ -0.7) as compared to 0.001 (\fe $\simeq$ -1.3). 
This would go in favour of a metal abundance slightly higher than measured
by FJ93.

There is the possibility that the FJ93 scale may
be underestimating cluster metallicities.  TAAT97 compared it  with abundances 
based on DDO photometry and found the FJ93 values systematically low.
Another example may be the couple of clusters examined by Gratton \& Contarini
(1994): they observed two giants in each cluster at high-resolution and high 
S/N (R=30,000, S/N $\simeq$ 100) and
found for NGC2243 and Mel66 the values [Fe/H]=--0.48 and --0.38
respectively, to be compared with [Fe/H]=--0.56 and --0.51 (FJ93).

FJ93 emphasized the fact that the actual value derived for the cluster
metallicity from their spectra is strongly dependent on the adopted reddening:
the $\pm$0.2 mag error on reddening in Janes (1991) allows for a formal
uncertainty of $\pm$0.3 dex in metallicity. They also find marginal support
for a \ebv ~value on the higher side, hence for a metal abundance slightly
higher than the [Fe/H]=--0.97 they give.
This goes in the same direction suggested by our comparisons, even if we do not
find any convincing evidence for a larger reddening.
Finally, we have identified the four stars studied by FJ93 (N$_{CJ}$ = 50,
406, 413, 415a, which correspond to N$_{our}$ = 50, 67, 20, 51 respectively), 
to check if they may be influenced by the differential reddening we found; but
we consider it quite improbable, since all the four objects are within 1
arcmin from the centre of the cluster.

No conclusive word can be said on Be21 metallicity, which would
instead be important to know with high precision, since it
could represent the lowest value for the open clusters in our Galaxy.
A decisive answer would come from high resolution spectroscopy coupled with
fine abundance analysis on the four stars examined by FJ93,
already known to be cluster members.

\bigskip\bigskip\noindent
ACKNOWLEDGEMENTS

\noindent
We warmly thank A. Chieffi, M. Limongi and, specially, O. Straniero 
for having not only distributed the new FRANEC tracks in advance of 
publication, but even in format suitable for our purposes.
We also thank J.C. Mermilliod for kindly making available his
invaluable BDA open clusters database and for useful comments.
We are grateful to the referee (Bruce Twarog) for his comments, extremely 
useful both to improve the clarity of the paper and for future applications.
The bulk of the numerical code for CMD simulations has been provided by Laura
Greggio. This research has made use of the Simbad database, operated at CDS,
Strasbourg, France.

\end{document}